\documentclass[aps,prl,twocolumn,showpacs,superscriptaddress,amssymb]{revtex4}

\usepackage{graphicx}
\usepackage{dcolumn}
\usepackage{textcomp}
\usepackage{amsmath}
\usepackage{hyperref}
\usepackage{natbib}
\usepackage{bm}

\begin{document}

\title{Deterministic nano-assembly of a coupled quantum emitter - photonic crystal cavity system}

\author{T. van der Sar$^{1}$, J. Hagemeier$^2$, W. Pfaff$^1$, E.C. Heeres$^3$, S.M. Thon$^2$, H. Kim$^{2,x}$, P.M. Petroff$^{4,5}$, T.H. Oosterkamp$^3$, D. Bouwmeester$^{2,3}$, and R. Hanson$^1$}

\affiliation{{$^1$Kavli Institute of Nanoscience, Delft University of Technology, P.O. Box 5046, 2600 GA Delft, The Netherlands.}
\newline{$^2$Department of Physics, University of California Santa Barbara, Santa Barbara, California 93106, USA.}
\newline{$^3$Leiden Institute of Physics, Leiden University, Niels Bohrweg 2, 2333 CA Leiden, The Netherlands.}
\newline{$^4$Department of Materials, University of California Santa Barbara, Santa Barbara, California 93106, USA.}
\newline{$^5$Department of ECE, University of California Santa Barbara, Santa Barbara, California 93106 USA.}
\newline{$^x$Current address: Department of ECE, IREAP, University of Maryland, College Park, Maryland 20742, USA.}}

\date{\today}

\begin{abstract}
Controlling the interaction of a single quantum emitter with its environment is a key challenge in quantum optics. Here, we demonstrate deterministic coupling of single Nitrogen-vacancy (NV) centers to high-quality photonic crystal cavities. We pre-select single NV centers and position their 50 nm-sized host nanocrystals into the mode maximum of photonic crystal S1 cavities with few-nanometer accuracy. The coupling results in a strong enhancement of NV center emission at the cavity wavelength.
\end{abstract}

\maketitle

Spontaneous light emission can be controlled by enhancing or suppressing the vacuum fluctuations of the electromagnetic field at the location of the light source~\cite{Mandel}. When placed into highly confined optical fields, such as those created in optical cavities or plasmonic structures, the optical properties of single quantum emitters can change drastically ~\cite{Novotny,Noda,Kuhn,Schietinger}. In particular, photonic crystal (PC) cavities show high quality factors combined with an extremely small mode volume ~\cite{Akahane}. It is challenging however to efficiently couple single photon sources to a PC cavity because the emitter has to be positioned in the localized optical mode, which is confined to an extremely small volume with a size of about a wavelength ~\cite{Hennessy,Rivoire,Thon}. 

Nitrogen-vacancy (NV) centers in diamond are promising candidates for application as solid state quantum bits ~\cite{Hanson,Neumann,Togan}. Long distance entanglement between NV centers enabling quantum repeater protocols may be achieved by two-photon quantum interference ~\cite{Barrett,Childress}, but is hindered by the NV centers' weak coherent photon emission rate. Being able to control and reshape the emission spectrum of a single NV center is therefore not only of fundamental interest, but could also have potential applications in solid state quantum information processing ~\cite{Jiang}.

Fabrication of high-quality PC cavities in diamond would be a natural way to control the emission properties of embedded NV centers, but this is challenging because of the difficulties in growing and etching diamond single-crystal thin films ~\cite{Greentree,Wang}. An alternative, hybrid approach is to position a diamond nanocrystal with a single NV center near a PC cavity of a different material ~\cite{Barth,Englund,Wolters,Stewart}. Because of the small size of such a crystal, the NV center can be placed in the highly confined optical mode where coupling can be efficient.

Here, we demonstrate the deterministic nano-assembly of coupled single NV center - PC cavity systems by positioning $\sim$50 nm sized diamond nanocrystals into gallium phosphide S1 cavities located on a different chip. The S1 cavity offers unique advantages over the well-studied L3
cavity ~\cite{Englund,Wolters}. Whereas in the L3 cavity the mode maximum is confined within the dielectric material, the mode maximum of the S1 cavity is localized in the air holes
surrounding the cavity, making it accessible for coupling to external emitters. We are able to pick up and place a pre-selected diamond nanocrystal exactly into the mode maximum of a PC cavity, due to the versatility of our nanopositioning method. The coupling of a single NV center to a PC cavity is evidenced by a Purcell-enhanced spontaneous emission rate at the cavity resonance
frequency.

Our samples were grown by molecular beam epitaxy on a (100) gallium phosphide (GaP) wafer, with a 1 $\mu$m sacrificial layer of Al$_{0.75}$Ga$_{0.25}$P, and a 120 nm GaP membrane layer. Photonic crystal cavities were defined by e-beam lithography, followed by inductively coupled plasma etching to transfer the resist pattern into the substrate. An HF chemical wet etch selectively undercuts the sacrificial layer, leaving a free-standing membrane. Structure parameters were optimized using Finite-Difference-Time-Domain (FDTD) simulations. The lowest energy mode of these cavities generally has the highest quality factor, up to 3,800 for S1 cavities, to our knowledge the highest reported value for a S1 photonic crystal cavity at these wavelengths.

\begin{figure}[ht!]
\begin{center}
\includegraphics[width=8.5cm]{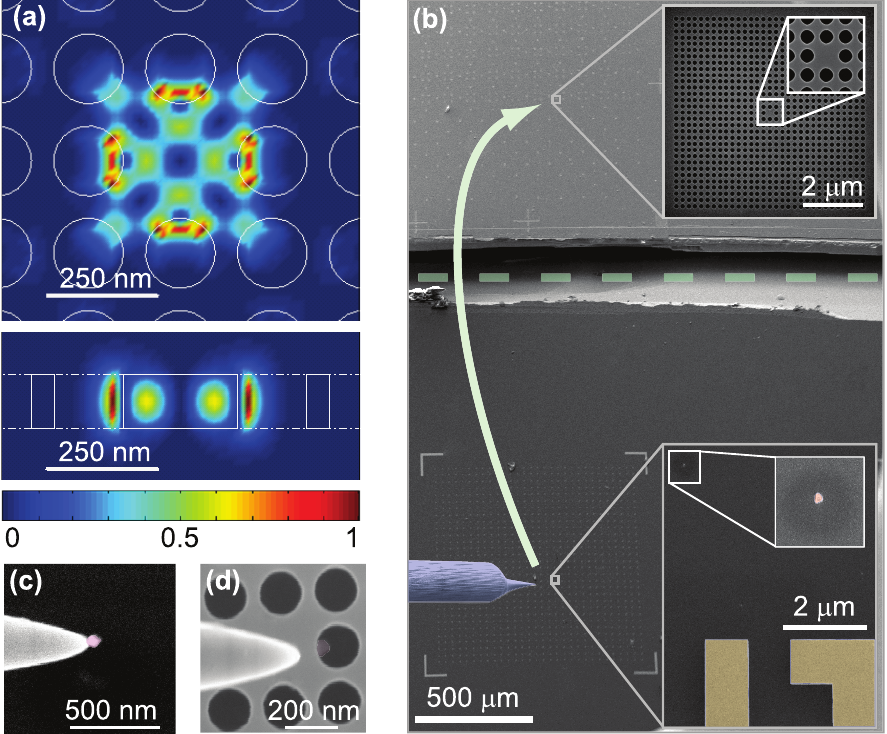}
\end{center}
\vspace*{-.5cm}
\caption{(color online) Nano-assembling a coupled single NV center -  photonic crystal cavity system. (a) Top view and cross section through the center of an S1 cavity showing the simulated intensity profile of the lowest energy mode. (b) SEM overview of the nanopositioning process. From the chip with nanocrystals (below the green dashed line), a nanocrystal that contains a single NV center is located using reference markers (lower inset), and subsequently picked up using a sharp tip (blue) and transferred to a chip containing photonic crystals (above the green dashed line). Top inset: close up of an S1 cavity. (c) SEM picture of a nanocrystal on the tip. (d) SEM picture showing a single NV containing nanocrystal that has been positioned into a hole of an S1 cavity. In b-d, colour shades were added to the tip, the gold marker and the nanocrystals for clarity.}
\vspace*{-.3cm}
\label{fig:1}
\end{figure}

We assemble a coupled single NV center - PC cavity system by nanopositioning a diamond nanocrystal into an air hole surrounding the cavity~\cite{vanderSar}, where the S1 cavity has its maximum field intensity (Fig. 1a). First, a home-built confocal microscope is used to locate and identify single NV centers on a chip that contains a sparse dispersion of diamond nanocrystals. We then select NV centers that do not show switching from the negatively charged state to the neutral state ~\cite{Bradac}. By accurately measuring the position of a candidate NV center with respect to a gold marker (fabricated using electron beam lithography), we can identify the host nanocrystal inside a scanning electron microscope (SEM) (Fig. 1b). This SEM contains a home-built nanomanipulator with a sharp tungsten tip that is mounted on a piezoelectric stage with sub-nanometer positioning accuracy ~\cite{Heeres}. With this tip we can pick up a nanocrystal (Fig. 1c), and position it in an arbitrary new location on a different chip. Here, we position a nanocrystal that contains a single NV center into the hole of an S1 cavity (Fig. 1d) ~\cite{Movies}.

When an NV center is successfully coupled to the cavity, it will show an enhanced spontaneous emission rate at the cavity wavelength by means of the Purcell effect~\cite{Mandel}. Compared to the spectrum of the NV center before positioning, the spectrum of the coupled system shows a sharp Fano-shaped resonance at the cavity frequency (Fig. 2a) ~\cite{Barclay,Galli,ZPL1}.

To verify that the spectrum originates from a single NV center we perform a second-order auto-correlation measurement on photon detection times. After correcting for background luminescence caused by substrate photoluminescence (PL), we find an antibunching dip below 0.5 (Fig. 2b) ~\cite{Kurtsiefer,ZPL}. The presence of a single NV center is also confirmed by the level of the observed PL. This in agreement with our previous studies ~\cite{vanderSar} where we found that single NV centers are not affected by the nanopositioning.

To verify that the NV center is coupled to the cavity, we measure the spectrum of the NV center in the cavity (middle panel Fig. 2a) as a function of optical excitation power. By fitting this data, (inset middle panel Fig. 2a), we obtain the amplitude of the cavity resonance and the zero-phonon line (ZPL) as a function of optical excitation power (Fig. 2c). In addition to a linear component caused by background luminescence coupling to the cavity, we observe the same saturation behavior for the cavity resonance as for the NV center's zero-phonon line. This in contrast to the purely linear increase of background luminescence measured in cavities without NV centers (data not shown). We conclude that the cavity is being fed by single photons from the NV center, as the finite optical lifetime of the NV center puts an upper limit on the photon emission rate ~\cite{AdditionalData}.

\begin{figure}[ht!]
\begin{center}
\includegraphics[width=8.5cm]{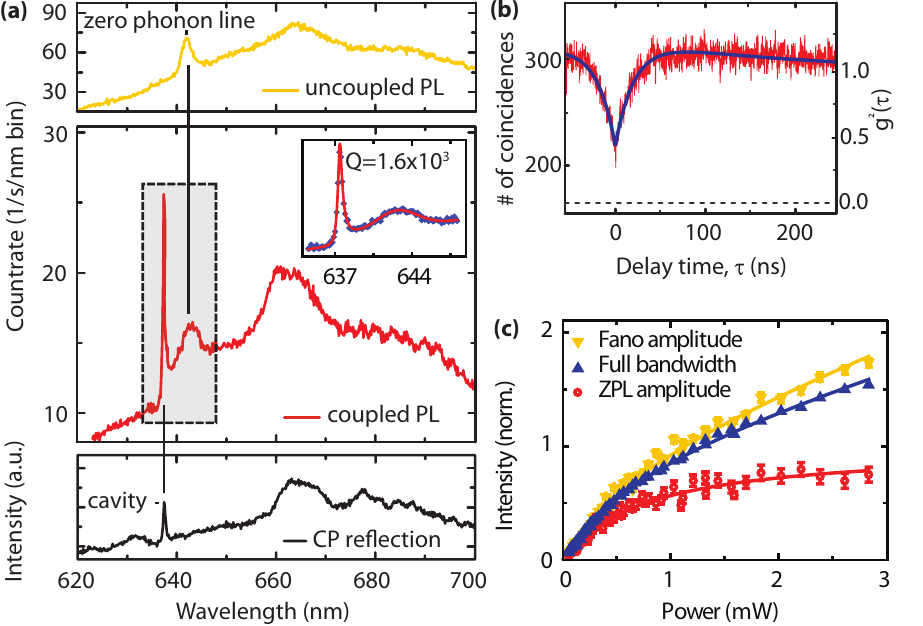}
\end{center}
\vspace*{-.5cm}
\caption{(color online) Characterization of a coupled NV center-cavity system. (a) Top panel: PL spectrum of a NV center before it was positioned into the nearest air hole of a S1 cavity (see Fig. 1d). Excitation: 300 $\mu$W at 532 nm. Middle panel: Spectrum of the coupled NV center - cavity system. A Fano resonance is observed at the cavity frequency close to the zero phonon line. Inset: Close up of the ZPL and the cavity line. The red line is a fit using a linear background plus a Gaussian and a Fano lineshape to model the shape of the ZPL and cavity line respectively. Excitation: 500$\mu$W at 568 nm (568 nm laser light induces less background PL than 532 nm). Bottom panel: CP reflection measurement showing the cavity resonance. All measurements were done at room temperature. (b) Measurement of the second-order auto-correlation function $g^2(\tau)$ of emission within a 615-700 nm bandwidth. Left axis: number of coincidences. Right axis: $g^2(\tau)$, corrected for background contribution estimated from APD signals measured next to the NV center (indicated by the dashed line) ~\cite{ZPL}. Excitation: 300 $\mu$W at 568 nm (c) Power dependence of the ZPL and Fano amplitude (obtained by a fit as in Fig. 2a), and of the integrated spectrum. Solid lines are fits with $y=y_{\infty}\frac{P}{P+P_{sat}}+aP$, where $P$ is the excitation power, and $y_{\infty}$ , $P_{sat}$ and $a$ are fit parameters. From the fit of the ZPL amplitude, which we assume to contain no background contribution ($a=0$), we extract $P_{sat}$ = 770 $\mu$W. The other curves were fit using this value. All curves are normalized to their respective $y_{\infty}$, so that the difference with the ZPL curve indicates the relative background contribution.}
\vspace*{-.3cm}
\label{fig:2}
\end{figure}

The Purcell-enhancement of the emission at the cavity wavelength can be estimated from the spectrum and the relative detection efficiencies of cavity emission and NV center emission. After subtracting the linear background contribution from the spectrum in Fig. 2a (middle panel), we find that coupling to the cavity has increased the relative intensity of detected NV emission at the cavity wavelength by a factor 4. This underestimates the actual Purcell enhancement because far-field detection of light emitted by an S1 cavity is relatively inefficient due to a large mode mismatch between the cavity and detection channel ~\cite{Badolato}. By comparing the detection efficiency of S1 cavity emission (obtained by calculating its far-field emission profile following the method described in \cite{Vuckovic2002}) to the detection efficiency of directly detected NV center emission, we estimate the actual Purcell enhancement to be 25. For comparison, the calculated Purcell enhancement for an optimally oriented NV center is 100. By adding efficient on-chip PC outcoupling structures or by optimizing the far-field emission profile of the cavity ~\cite{Tran,Portalupi}, the detection efficiency can be increased, allowing the full cavity-enhanced emission to be exploited.

In conclusion, we have demonstrated the coupling of single NV centers to S1 PC cavities by positioning 50 nm-sized diamond nanocrystals exactly into the cavity mode maximum. By pre-selecting nanocrystals with single NV centers that have rare but highly desirable optical properties such as a lifetime limited linewidth ~\cite{Shen}, our method could be applied to reshape the NV center spectrum for a more efficient emission of coherent photons. We envision that nano-assembled, coupled NV center-cavity systems could serve as building blocks for quantum optical engineering, potentially leading to the first observation of entanglement between distant NV centers.

This work is supported by stichting voor Fundamenteel Onderzoek der Materie (FOM) and the Nederlandse Organisatie voor Wetenschappelijk Onderzoek (NWO). T.H.O. acknowledges support from Technologiestichting STW and from an ERC Starting Grant. D. B. acknowledges support from NSF grant 0901886 and Marie Curie EXT-CT-2006-042580. J. H. acknowledges support from a U. S. Department of Education GAANN grant.


\begin{thebibliography}{}
\bibitem{ZPL1}
The zero-phonon line is at 642 nm, a larger wavelength than the 637 nm usually found for NV centers in bulk diamond. This is due to high strain commonly present in diamond nanocrystals.
\bibitem{ZPL}
Background contribution was measured on the photonic crystal and not on the cavity. However, background is less on the photonic crystal because there is less substrate material (due to the holes). We therefore underestimate the background contribution.
\end{thebibliography}
\end{document}